\documentclass[12pt]{article}
\usepackage{epsfig}
\usepackage{euscript}

\def\beq{\begin{equation}}
\def\eeq{\end{equation}}
\def\bea{\begin{eqnarray}}
\def\eea{\end{eqnarray}}
\def\bq{\begin{quote}}
\def\eq{\end{quote}}

\newcommand{\rr}{\mbox{\boldmath $r$}}
\newcommand{\rrn}{\mbox{$r$}}

\def\lfrestriction#1{\lower.25ex\hbox{\Big|}_{#1}}  

\def\gappeq{\mathrel{\rlap {\raise.5ex\hbox{$>$}}
{\lower.5ex\hbox{$\sim$}}}}

\def\lappeq{\mathrel{\rlap{\raise.5ex\hbox{$<$}}
{\lower.5ex\hbox{$\sim$}}}}

\def\Toprel#1\over#2{\mathrel{\mathop{#2}\limits^{#1}}}

\begin{document}


\begin{center}
\begin{Large}
{\bf Extraction of the Skewing factor from the DIS/DVCS ratio.} 
\\ 
\end{Large}

\vspace*{1cm}
L. Favart$^{a}$, M.V.T.~Machado$^{b}$ and L.~Schoeffel$^{c}$\\
\vspace{1.0cm}
{$^{a}$ \rm I.I.H.E., Universit\'e Libre de Bruxelles, 1050 Brussels, Belgium}\\\vspace{0.3cm}
$^{b}$ \rm High Energy Physics Phenomenology Group, GFPAE,  IF-UFRGS \\
Caixa Postal 15051, CEP 91501-970, Porto Alegre, RS, Brazil\\
\vspace{0.3cm}
$^{c}$ \rm CE Saclay, DAPNIA-SPP, 91191, Gif-sur-Yvette,
France\\
\end{center}

\begin{abstract}
 The skewing factor, defined as the ratio of the imaginary parts of the
amplitudes ${Im \,{\cal A}(\gamma^*\, p \to \gamma^*\, p)}/{Im \,{\cal
A}(\gamma^*\, p \to \gamma\, p)}$ is extracted for the first time from the
data using recent Deeply Virtual Compton Scattering (DVCS) and the
inclusive inelastic cross section measurements at DESY-HERA. The results
values are compared to theoretical predictions for NLO QCD and the colour
dipole approach. 
\end{abstract}

\section*{Introduction}

It is well known that the cross section of hard scattering processes can be
described as the convolution of parton distributions (PDFs) and the cross
sections of hard subprocesses computed at parton level using perturbative
QCD. The usual PDFs, obtained from inclusive experimental data, is the
diagonal element of an operator in the Wilson operator product expansion
(OPE). On the other hand, there is a set of exclusive reactions which are
described by the off-diagonal elements of the density matrix, where the
momentum, helicity or charge of the outgoing target are not the same as
those of the corresponding incident particle. Examples of such reactions
are the virtual photon Compton scattering (DVCS) ($\gamma^* p \rightarrow
\gamma p$) and the vector meson electroproduction ($\gamma^* p \rightarrow
V p$). In these cases, the difference with the inclusive case is the
longitudinal components of the incoming and outgoing proton momentum,
which depend on the photon virtuality $Q^2$ and the $\gamma^* p $ center
of mass energy $W$.

\begin{figure}[t]
 \begin{picture}(100,100)(0,0)
  \put(5.0,0.0){\epsfig{figure=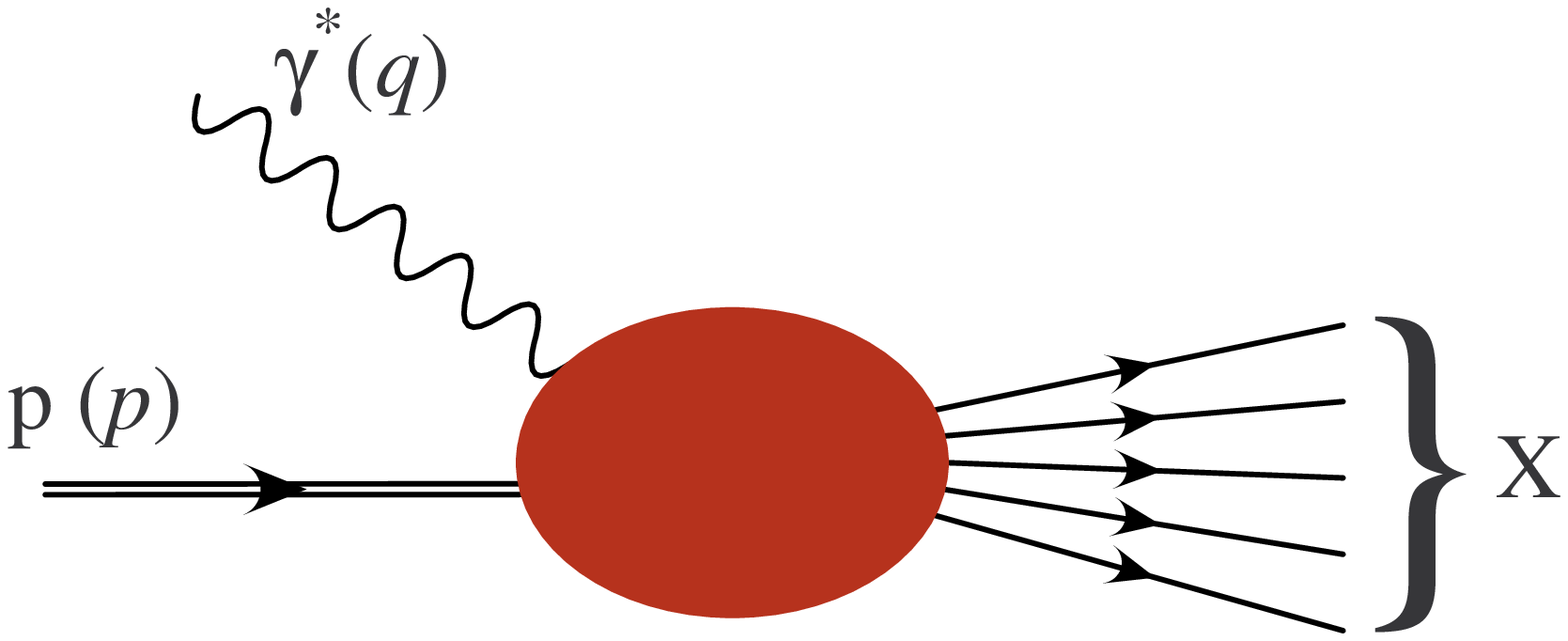,width=0.45\textwidth}}
  \put(205.0,10.0){\epsfig{figure=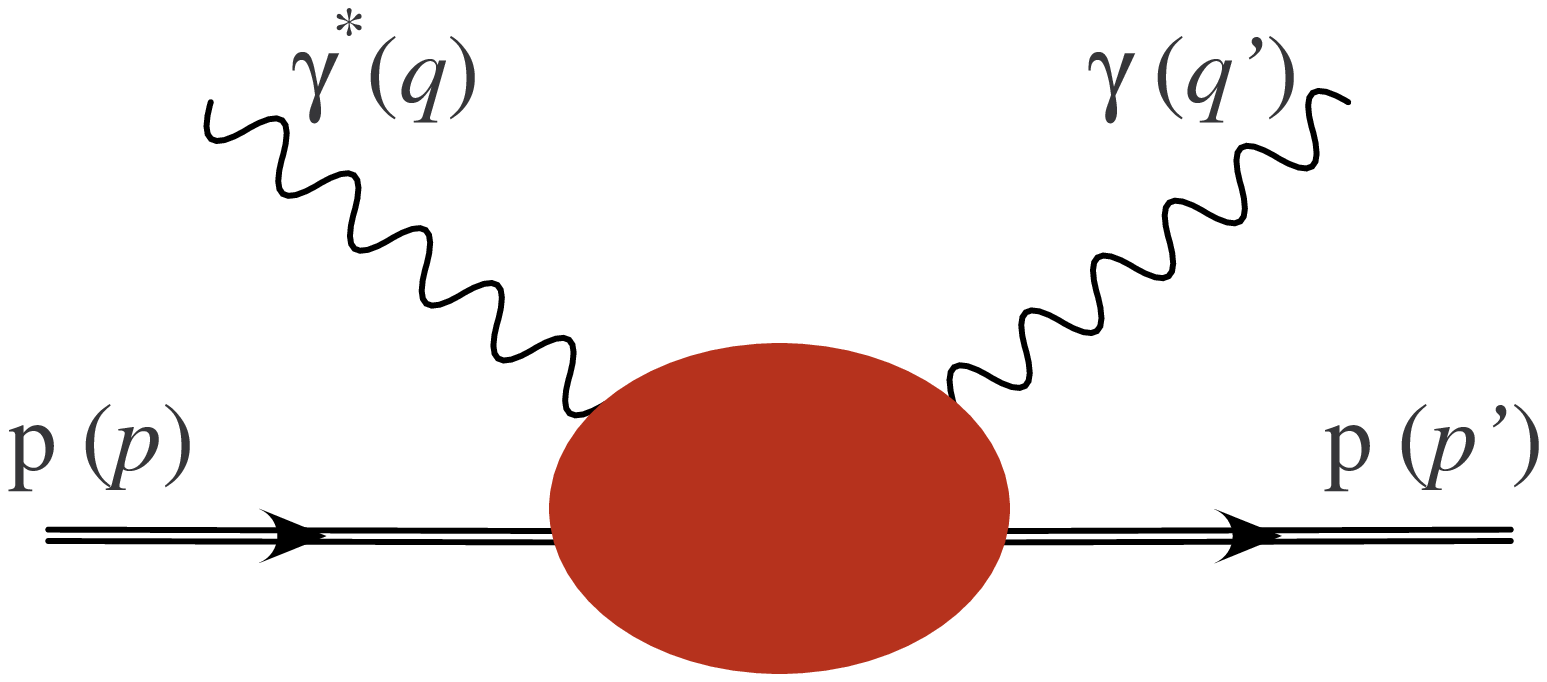,width=0.45\textwidth}}
  \put(5.0,0){a)}
  \put(205.0,0){b)}
 \end{picture}
\caption{Amplitude level diagrams for DIS (a) and DVCS (b)}
\label{fig:diags}
\end{figure}

Precision data are becoming available for hard scattering processes whose
description requires knowledge of these off-diagonal (or skewed) parton
distributions. The exclusive diffractive DVCS process at large $Q^2$
provides a comparatively clean procedure for extracting information on the
gluons within the proton in a non-forward kinematic case. In recent years
the study of Deep Inelastic lepton-proton Scattering (DIS)
has produced detailed information of the proton structure in terms of
Parton Distribution Functions (PDFs). The optical theorem states that the
leading DIS process of a single virtual photon exchange can be viewed as
identical to the forward elastic scattering of a virtual photon from the 
proton. This scattering, known as the forward Compton scattering
process, is the same as a DIS process (see Fig.~\ref{fig:diags}-a) with
its mirrored reaction and relates the total cross section
$\sigma_{tot}(\gamma^* p \rightarrow X)$ with the imaginary part of the
forward amplitude $Im\, A(\gamma^* p \rightarrow \gamma^* p)$.

The forward Compton scattering
$\gamma^*(q) \, p(p) \rightarrow \gamma^*(q) \, p(p)$ can be generalized 
to the non forward case (at non-zero angles): 
$\gamma^*(q) \, p(p) \rightarrow \gamma^{(*)}(q_2) \, p(p_2)$ 
with $q_2\ne q$ and $p_2 \ne p$ and being experimentally accessible through the
so called Deeply Virtual Compton Scattering (DVCS) process, where a real
photon is observed in the final state (see Fig.~\ref{fig:diags}-b). In
analogy with DIS which relates the imaginary part of the forward Compton
amplitude and PDFs, in the DVCS process parton distributions are also used
to parameterise the non-perturbative part of the interaction in the
non-forward region. In this last case, the distributions have to be
calculated in terms of non-diagonal, rather than diagonal, PDFs. They are
commonly called Generalised Parton Distributions (GPDs).  They reduce to
the ordinary PDFs in the limit of zero skewedness, i.e., the forward
limit.
\\

Let us first summarize the main theoretical approaches describing 
the DVCS process.  
From the perturbative QCD point of view, GPDs have been studied
extensively and incorporate the partonic and distributional
behavior.  They are two-parton correlation functions that allow accessing
parton correlations within hadrons. For fixed skewedness, they are
continuous functions and cover two distinct regions, the DGLAP and the
ERBL regions, in which their evolution with the hard scale is driven by the
corresponding evolution equations. Moreover, the Lorentz structure of the
GPDs implies a polynomiality condition; namely, their $(n-1)$-th moments
are polynomials in the square of the skewedness of degree smaller than
$n/2$. The most promising approach for GPD parameterisations relies on 
setting them equal
to the forward PDFs using a forward model with suitably symmetrized input
GPDs in the ERBL region constructed in order to satisfy polynomiality for
the first two momenta~\cite{Freund:2002qf}. 
These theoretical studies reproduce satisfactorily the
main features of DVCS data~\cite{h1dvcs}. 
On the other hand, a simple and intuitive high
energy formalism has recently been used to describe 
DVCS~\cite{Favart:2003cu}. 
It is given by a
colour dipole picture, which allows extending in the studies of
saturation physics at low virtualities of the incoming photon. This is an
advantage in comparison with the pQCD formalism which is limited by the
initial scale for QCD evolution to the order of $Q_0 \gappeq 1$ GeV. 
Additionally,
the roles played by QCD evolution and skewing effects have been recently
shown in the scope of this formalism. 
The off-diagonal effect has been investigated
through a simple phenomenological parameterisation. The overall picture is
in good agreement with the experimental data~\cite{Favart:2004uv}. 
\\

 In this work, we will
address the question of the extraction of the skewing correction using the
available data on DIS and DVCS. This can be performed by computing the ratio
of the imaginary parts of the amplitudes for these reactions. The skewing
factor $R$ will depend on the experimental measurements and therefore
possess
a small degree of model dependence. We therefore have the attractive
possibility in determining its dependence upon $Q^2$ virtuality and upon
$W$ energy. This knowledge will enable us to gain insight into the skewedness
correction to the non-forward observables. Namely, the non-forward cross
section and distributions can be easily obtained from the forward ones by
simply multiplying them by the skewing factor. This has direct implications
in computing the exclusive meson and heavy boson production via the
forward scattering amplitude, which is better constrained from the
inclusive measurements. 
\\

The paper is organized as follows; in the next
section, the skewing factor $R$ is defined and extracted from experimental
DESY-HERA results on DIS and DVCS. In section~\ref{sec:pred}
we perform a comparison of the
skewing factor with two different theoretical approaches, the perturbative
NLO QCD calculation and the colour dipole formalism.  A qualitative
analysis of the extracted skewing factor is then performed in a colour dipole
approach. In the last section we summarize the main results and discussions.

\section{Definition and extraction of the skewing factor $R$}

Lets define a basic quantity giving an overall measurement of the skewing
properties, which includes both the non-forward kinematics and the
non-diagonal effects. Namely, we set the ratio between the imaginary parts
of the DIS and DVCS (forward) scattering amplitudes at zero momentum
transfer:
\beq
R \equiv \frac{Im\, {\cal  A}\,(\gamma^*+p \to \gamma^* +p)\lfrestriction{t=0}}
{Im \,{\cal A}\,(\gamma^*+p \to \gamma +p)\lfrestriction{t=0}} \ ,
\label{R_def}
\eeq
where $t$ is the square of the four-momentum exchanged at the proton vertex.
\\

The scattering amplitude for the DIS process can be directly obtained from the
DIS cross section and experimentally measured at DESY-HERA, that is 
$ Im\, {\cal A}(\gamma^* p \rightarrow \gamma^* p) \sim 
\sigma_{tot}(\gamma^* p \rightarrow X)$. 
In fact, the DIS amplitude can be written down using the
usual pQCD fits for the proton structure function, $\sigma_{tot}^{\gamma^*
p}=(4\pi^2\alpha/Q^2)\,F_{2}^p(x,Q^2)$.  The DVCS scattering amplitude can
be obtained from the recent measurements on the Deeply Virtual Compton
Scattering cross section. In this case, the $t$ dependence of the
amplitude can be assumed to be factorised out and parameterised as
an exponential, $\propto \exp (-b|t|)$, and the total DVCS cross section can
be related to the corresponding amplitude at $t=0$; namely,
\begin{eqnarray}
 \sigma(\gamma^*\,p\rightarrow \gamma \,p) 
 & = & \frac{\left[Im \,{\cal A}\,(\gamma^*p \to \gamma 
   p)\lfrestriction{t=0}\right]^2}{16\pi\,b}\,, \label{dvcs_xs}
\end{eqnarray} where $b$ is the $t$ slope parameter referred to above. 
The expression above can
be corrected by taking into account the contribution from the real part of
the amplitude by multiplying Eq. (\ref{dvcs_xs}) by a factor $(1+\eta^2)$,
where $\eta$ is the ratio of the real to imaginary part of the DVCS amplitude.
The typical contribution for the kinematic window available at
HERA is of the order of 20 \%. In particular, to a good approximation, $\eta$ can
be calculated using dispersion relations to $\eta \simeq \tan (\pi\lambda
/2)$, where $\lambda=\lambda (Q^2)$ is the effective power of the
Bjorken $x$ dependence of the imaginary part of the amplitude.  
For an estimate of the real part
contribution we use the effective power for the inclusive deep inelastic
reaction taken from Ref. \cite{h1lambda}.

Considering the calculation discussed above, we can rewrite the skewing
factor as a function of the total cross sections for DIS and DVCS.
The factor reads as, 
\beq
 R = \frac{\sigma(\gamma^* \,p\rightarrow X)\, \sqrt{(1+\eta^2)}}
 {4\,\sqrt{\pi b\,\sigma(\gamma^*\,p\rightarrow \gamma \,p)}} \,
=\frac{\sqrt{\pi^3}\,\alpha}{Q^2} \, \frac{F_2^p(x,Q^2)\, \sqrt{(1+\eta^2)}}
 {\sqrt{b\,\sigma(\gamma^* \,p\rightarrow \gamma \,p)}} \,.  
\label{R_def_ap} 
\eeq 
Main theoretical uncertainties come from the $b$ slope and from the
estimate of the real part contribution. In our further calculation, one
uses the $b$ value extracted from the recent measurements of the
$t$-dependence of DVCS data.  
\\

\begin{figure}
 \begin{center}
 \epsfig{figure=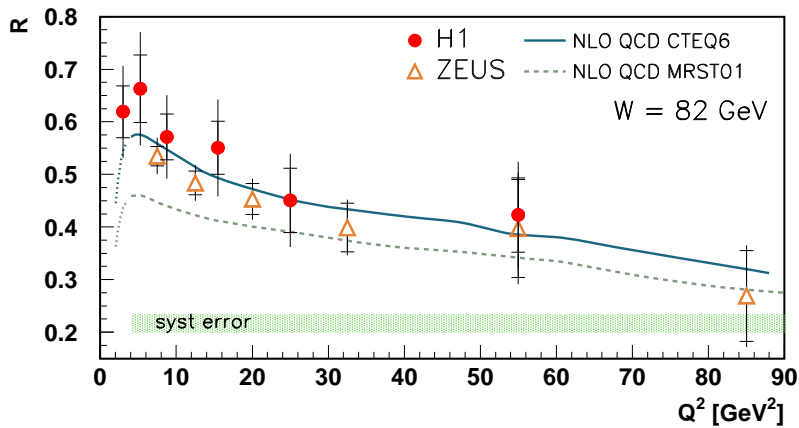,width=0.78\textwidth}
 \end{center}
 \vspace*{-0.5cm}
 \caption{\sl The skewing factor $R$ as a function of $Q^2$ at $W~=~82$~GeV.
  The points correspond to the $R$ extraction applied to DVCS measurements 
   of H1
  (bullets) and ZEUS (triangles).  The curves represent $R$ extracted from
  Freund {\it et al.} prediction based on MRST 2001 and CTEQ6 PDFs.
}
 \label{fig:r_qcd_q2}
\end{figure}

\begin{figure}
 \begin{center}
  \epsfig{figure=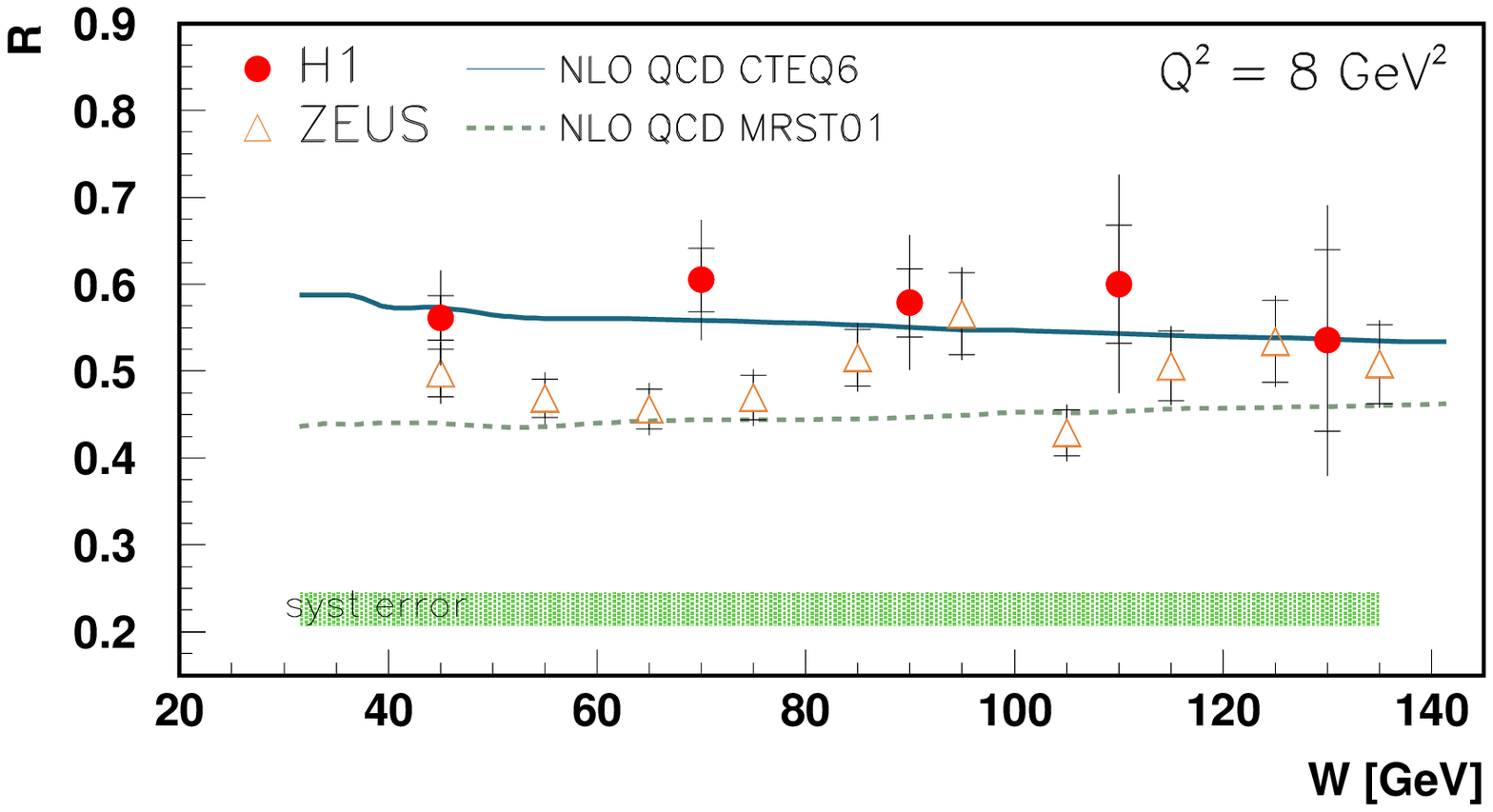,width=0.78\textwidth}
 \end{center}
 \vspace*{-0.5cm}
 \caption{\sl The skewing factor $R$ as a function of $W$ at $Q^2=8$~GeV$^2$.
  (see caption of Fig.~\ref{fig:r_qcd_q2}).
}
 \label{fig:r_qcd_w}
\end{figure}

To extract the
factor $R$ factor, Eq.~(\ref{R_def_ap}), we use recent DVCS measurements
at HERA~\cite{h1dvcs,zeusdvcs} and the DIS cross cross section is obtained
using the pQCD fits of the $F_2^p$ structure function from Ref.
\cite{h1f297data}. The factor $R$ is shown as a function of $Q^2$ in
Fig.~\ref{fig:r_qcd_q2} and as a function of energy $W$ in
Fig.~\ref{fig:r_qcd_w}. 
The corresponding values are given in tables \ref{tab:res1} and \ref{tab:res2}.
The inner error bars represent the statistical
error. The full error bars is the quadrature sum of the statistical,
systematic and normalisation (precision of the $b$ measurement of H1)
uncertainties\footnote{important contribution to systematic effects in
the H1 measurement are not considered in the ZEUS measurement (e.g.\ the
error on the Bethe-Heitler subtraction which dominates the full
systematic error at large $W$ values)}. 
A $Q^2$ dependence is observed, decreasing from $R\simeq
0.7$ for $Q^2\simeq2$ GeV$^2$ down to $R\simeq 0.3$ for $Q^2\simeq 85$
GeV$^2$. Qualitatively, the data seem to be consistent with a dependence
like $R\propto 1/\log Q^2$. An almost flat $W$ dependence is observed
within the present precision. This feature can be easily understood by
inspecting Eq. (\ref{R_def_ap}), since the $W$ dependence of both the DIS and
DVCS cross section is power-like having a proportional effective power.
Namely, $\sigma_{\mathrm{DIS}}\propto W^{2\lambda}$ and
$\sigma_{\mathrm{DVCS}}\propto W^{4\lambda}$. The mean value $R\simeq 0.5$
is consistent with its early theoretical estimates using the aligned jet
model \cite{Frankfurt:1997at} and the colour dipole saturation model
\cite{Favart:2003cu}.

\begin{table}[htbp]
\begin{tabular}{ll}
\begin{tabular}{|c|lcc|}
 \hline 
 $Q^2$  [GeV$^2$]& \multicolumn{3}{c|}{ $R$ using H1 data}\\
 \hline
 3    &$\; 0.62 $ &$ \pm 0.05 $ &$ \pm 0.07  $ \\
 5.25 &$\; 0.66 $ &$ \pm 0.06 $ &$ \pm 0.09  $ \\
 8.75 &$\; 0.57 $ &$ \pm 0.04 $ &$ \pm 0.07  $ \\
 15.5 &$\; 0.55 $ &$ \pm 0.05 $ &$ \pm 0.08  $ \\
 25   &$\; 0.45 $ &$ \pm 0.06 $ &$ \pm 0.06  $ \\
 55   &$\; 0.42 $ &$ \pm 0.07 $ &$ \pm 0.07  $ \\
 \hline 
\end{tabular}
&

\begin{tabular}{|c|lcc|}
 \hline
 $Q^2$ [GeV$^2$] & \multicolumn{3}{c|}{ $R$ using ZEUS data}\\
 \hline
 7.5   & 0.53 &$ \pm 0.02 $ &$ \pm 0.03  $ \\
 12.5  & 0.48 &$ \pm 0.02 $ &$ \pm 0.02  $ \\
 20    & 0.45 &$ \pm 0.03 $ &$ \pm 0.02  $ \\
 32.5  & 0.40 &$ \pm 0.05 $ &$ \pm 0.03  $ \\
 55    & 0.41 &$ \pm 0.09 $ &$ \pm 0.05  $ \\
 85    & 0.28 &$ \pm 0.09 $ &$ \pm 0.04  $ \\
 \hline 
\end{tabular}
\end{tabular}
\caption{\sl The skewing factor $R$ for different $Q^2$ values at
$W~=~82$~GeV according to the DVCS cross section measurement of H1
(left table) and ZEUS (right table).
}
\label{tab:res1}
\end{table}

\begin{table}[htbp]
\begin{tabular}{ll}
\begin{tabular}{|c|lcc|}
 \hline 
 $W$ [GeV] & \multicolumn{3}{c|}{ $R$ using H1 data}\\
 \hline
 45   & 0.56 &$ \pm 0.03 $ &$ \pm 0.05  $ \\
 70   & 0.60 &$ \pm 0.04 $ &$ \pm 0.06  $ \\
 90   & 0.59 &$ \pm 0.04 $ &$ \pm 0.07  $ \\
 110  & 0.60 &$ \pm 0.07 $ &$ \pm 0.10  $ \\
 130  & 0.53 &$ \pm 0.10 $ &$ \pm 0.11  $ \\

 \hline 
\end{tabular}
&

\begin{tabular}{|c|lcc|}
 \hline
 $W$ [GeV] & \multicolumn{3}{c|}{ $R$ using ZEUS data}\\
 \hline
 45   & 0.50 &$ \pm 0.03 $ &$ \pm 0.02  $ \\
 55   & 0.47 &$ \pm 0.02 $ &$ \pm 0.02  $ \\
 65   & 0.46 &$ \pm 0.02 $ &$ \pm 0.02  $ \\
 75   & 0.47 &$ \pm 0.03 $ &$ \pm 0.02  $ \\
 85   & 0.52 &$ \pm 0.03 $ &$ \pm 0.02  $ \\
 95   & 0.57 &$ \pm 0.05 $ &$ \pm 0.03  $ \\
 105  & 0.43 &$ \pm 0.03 $ &$ \pm 0.02  $ \\
 115  & 0.51 &$ \pm 0.04 $ &$ \pm 0.02  $ \\
 125  & 0.53 &$ \pm 0.05 $ &$ \pm 0.02  $ \\
 135  & 0.51 &$ \pm 0.04 $ &$ \pm 0.02  $ \\
 \hline 
\end{tabular}
\end{tabular}
\caption{\sl The skewing factor $R$ for different $W$ values at
$Q^2=8$~GeV$^2$ according to the DVCS cross section measurement of H1
(left table) and ZEUS (right table).
}
\label{tab:res2}
\end{table}

\section{Theoretical Predictions}
\label{sec:pred}

 In this section we compare the results extracted from experimental data
to different theoretical predictions. First, we contrast them with the
perturbative QCD approach at NLO accuracy. This prediction is dependent on
the Generalised Parton Distributions (GPD), which have been parameterised
in Ref.~\cite{Freund:2002qf} and applied to describe the recent DVCS data
~\cite{Freund:2001hm,Freund:2001hd}. 
The second theoretical approach~\cite{Favart:2003cu,Favart:2004uv} is
given by the colour dipole approach, where the basic building blocks are 
the photon
wavefunctions and the dipole cross section. The non-forward kinematics is
encoded by the wavefunctions whereas the off-diagonal effects can be
built-in in the parameterisations for the dipole-target cross section.

\subsection{NLO QCD Predictions}

 The DVCS cross section has been calculated at NLO in perturbative QCD 
by Freund and McDermott~\cite{Freund:2001hm,Freund:2001hd}
using two different GPD
parameterisations~\cite{Freund:2002qf}.
The MRST2001 and CTEQ6 parameterisations of the PDFs
are used in the DGLAP region and polynomial form is used in the ERBL region,
ensuring a smooth continuation between the two regions.
Both the skewing ($\xi$) and the $Q^{2}$ dependence are generated dynamically.
The $t$ dependence is factorised out and assumed to be $e^{-b|t|}$. In its
original form, the approach above used a $Q^2$-dependent $b$ slope. We
have verified that the recent data on DVCS \cite{h1dvcs} can be reasonably
described using a fixed $b$ slope. 
\\

 Applying the same method presented here (i.e.\ using the same
$F_2^p$) for the inclusive cross section, the values obtained are shown 
as a function of $Q^2$ in Fig.~\ref{fig:r_qcd_q2}
and as a function of $W$ in Fig.~\ref{fig:r_qcd_w}. 
The value of $b=6.02\pm 0.35\pm 0.39$ GeV$^{-2}$ 
measured by H1 \cite{h1dvcs} has been used in all predictions. The error on 
the $b$ value  is not applied to theoretical predictions as it is 
already included
in the total error of the data points.
A good agreement between these predictions and the data points is found,
describing well the absolute value of $R$ and its kinematic dependences.
The NLO calculation presents no $W$ dependence.
The result normalisation is strongly dependent on the
choice for the gluon distribution and the deviations seem to be amplified
at lower $Q^2$. Concerning the behavior on $Q^2$, it seems 
that for $Q^2$ values below $\simeq 4$ GeV$^2$ the DGLAP evolution
starting at $Q^2_0 = 1$ GeV$^2$ has too little phase space to fully
generate the gluon distributions (dotted part of the curves in
Fig.~\ref{fig:r_qcd_q2}).
The underestimation of $R$ for virtualities below 4 GeV$^2$ may be 
due to a larger value for the DVCS cross section calculated at NLO where
the gluon distribution is underestimated as an effect of the too small 
phase space for evolution. 

\subsection{Colour Dipole Model Predictions}

In the colour dipole approach, the DIS (or DVCS) process can be seen as a
succession in time of three factorisable subprocesses: i)  the virtual
photon fluctuates in a quark-antiquark pair, ii) this colour dipole
interacts with the proton target, iii) the quark pair annihilates into a
virtual (or real) photon. The imaginary part of the DIS (or DVCS)
amplitude at $t=0$ is expressed in the simple way \cite{Favart:2003cu},
\begin{eqnarray} 
 {\cal I}m\, {\cal A}\,(W,Q_1,Q_2)  =  \sum_{T,L}\int \limits_0^1 dz \! 
 \int\limits_{0}^ {\infty} d^2\rr\, 
 \Psi_{T,L}^*(z,\,\rr,\,Q_1^2)\,\sigma_{dip}\,(\tilde{x},\rr)\Psi_{T,L}
 (z,\,\rr,\,Q_2^2)\label{dvcsdip}\,,
\end{eqnarray} 
where $\Psi(z,\rr,Q_{1,2})$ are the light cone photon wave functions for
transverse and longitudinal photons. The quantity $Q_1$ is the virtuality
of the incoming photon, whereas $Q_2$ is the virtuality of the outgoing
photon. 
In the DIS case, one has $Q_1^2=Q_2^2=Q^2$ and for DVCS,
$Q_1^2=Q^2$ and $Q_2^2=0$. 
The relative transverse quark pair (dipole) separation is labeled by 
$\rr$ whilst $z$ (respec.\ $1-z$) labels the quark (antiquark)
longitudinal momentum fraction.
\\

It should be noticed that the non-forward kinematics for DVCS is encoded in
the colour dipole approach through the different weight coming from the
photon wavefunctions in Eq. (\ref{dvcsdip}). The off-diagonal effects,
which affect the gluon and quark distributions in the pQCD approaches,
should be included in the parameterisation of the dipole cross section. At
the present stage of the development of the colour dipole formalism, we
have no accurate theoretical arguments on how to compute skewedness effects
from first principles. A consistent approach would be to compute the
scattering amplitude in the non-forward case, since the non-forward photon
wave function has been recently obtained in Ref.~\cite{Bartels:2003yj}.
In this case, the dipole cross section,
$\sigma_{dip}\,(x_1,x_2,\rr,\vec{\Delta })$, depends on the light cone
momenta $x_1$ and $x_2$ carried by the exchanged gluons, respectively, and
on the total transverse momentum transfer $\vec{\Delta}$. In this case,
additional information about the dependence upon $\vec{\Delta}$ is needed for
the QCD Pomeron and proton impact factor. The forward dipole cross section
is recovered at $x_1=x_2$ and $\vec{\Delta}=0$.
\\

In Ref. \cite{Favart:2004uv} an estimate of the skewedness for the dipole cross
section has been performed using the approximation of 
Ref.~\cite{Shuvaev:1999ce} where the ratios of off-diagonal to diagonal
parton distributions are computed. The behavior of those ratios are given
explicitly by:
\begin{eqnarray}
  R_{q,\,g}\,(Q^2)=\frac{2^{2\lambda + 3}}
   {\sqrt{\pi}}\,\frac{\Gamma\,\left(\lambda+ 
    \frac{5}{2}\right)}{\Gamma \,\left(\lambda+3+p \right)}\,,
 \label{eq:skew}
\end{eqnarray}
where $p=0$ for quarks and $p=1$ for gluons, whereas $\lambda$ is
the effective exponent of the parton distribution. The ratio is
larger for singlet quarks than for gluons. In the colour dipole picture, the
DVCS cross section is driven mainly by gluonic exchanges. In our numerical
computations, we use $\lambda=\lambda(Q^2)$ as obtained from the DVCS
scattering amplitude and the skewedness effect is given by multiplying the
original (without skewedness) total cross section by the factor
$R_g^2(Q^2)$. A different implementation of the skewedness correction is
to use the approximation $\tilde{x}=0.41\,x_{\mathrm{Bj}}$, which produces
the same effects as for $R_g$ at the presently covered kinematic range and
precision of the measurement. This different procedure is particularly
suitable for dipole cross sections which are parameterised by using a gluon
distribution, namely $\sigma_{dip}\propto \rr^2xg\,(x,c/\rr^2)$.

\begin{figure}
 \begin{center}
  \epsfig{figure=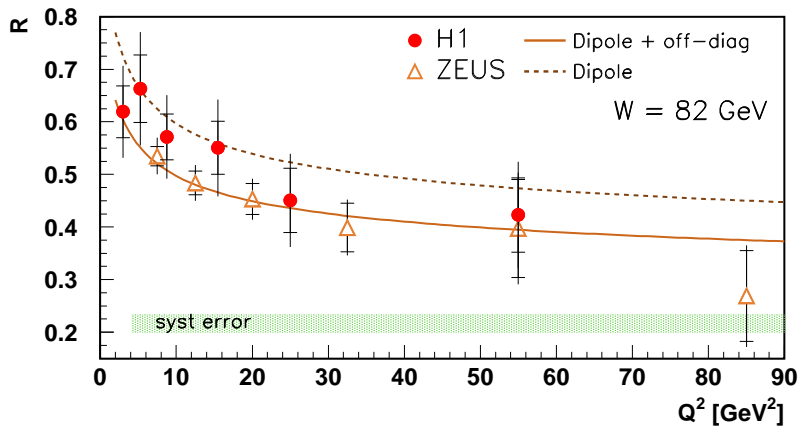,width=0.78\textwidth}
 \end{center}
 \vspace*{-0.5cm}
 \caption{\sl The skewing factor $R$ as a function of $Q^2$ at $W~=~82$~GeV.
  The curves represent the theoretical predictions of the colour dipole
  approach (see
  text) with (solid) and without (dashed) applying the off-diagonal
  skewing factor $R_g$.
}
 \label{fig:r_dip_q2}
\end{figure}

\begin{figure}
 \begin{center}
  \epsfig{figure=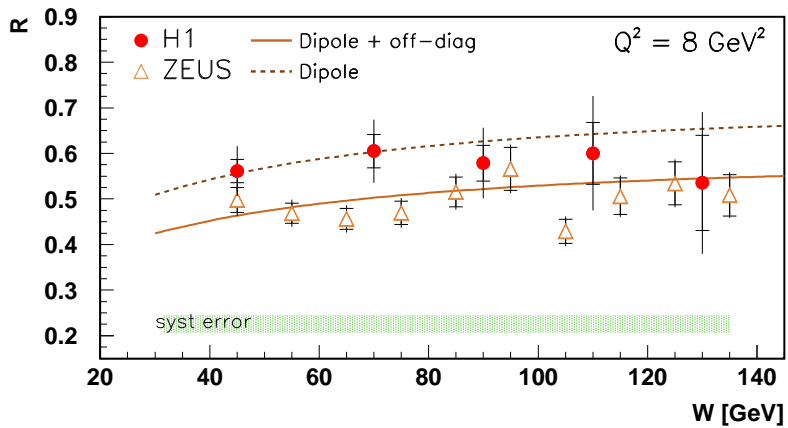,width=0.78\textwidth}
 \end{center}
 \vspace*{-0.5cm}
 \caption{\sl The skewing factor $R$ as a function of $W$ at $Q^2=8$~GeV$^2$.
  (see caption of Fig.~\ref{fig:r_dip_q2}).
}
 \label{fig:r_dip_w}
\end{figure}

Using this colour dipole prediction (with DGLAP evolution and
off-diagonal skewing factor $R_g$), one can again 
extract the $R$ factor, as shown as a function of $Q^2$ in 
Fig.~\ref{fig:r_dip_q2}
and as a function of $W$ in Fig.~\ref{fig:r_dip_w}. 
A good agreement with points extracted from the data is found. 
The curve obtained without applying the off-diagonal
skewing factor $R_g$ is also presented, enabling one to isolate
the two effects: the non-forward kinematic and the off-diagonal 
gluon distributions. En passant these results demonstrate that the
approximation of Eq.~\ref{eq:skew} is reliable within the present
precision. Furthermore, the off-diagonal effects contribute mostly for the
overall normalisation of the skewing factor whereas the behavior on $Q^2$
seems to be driven by the non-forward kinematic effect.
This prediction exhibits a moderate increasing of $R$ with $W$
which is not present in the NLO prediction.

Let us perform a qualitative analysis
in order to obtain a physical picture for the $W$ and $Q^2$ dependences of
the skewing factor. We take the colour dipole approach and the saturation
model for the dipole cross section ~\cite{Golec-Biernat:1999qd},
considering for simplicity the massless case. The final result should be
sufficiently independent of this particular choice for the dipole cross
section. One can use an approximated form for the dipole cross section from
the saturation model; $\sigma_{dip}=\sigma_0\,\rrn^2\, Q_s^2(x)/4$ (colour
transparency) for $\rrn^2 \leq 4/Q_s^2(x)$ and $\sigma_{dip}=\sigma_0$
(black disk limit) for $\rrn^2>4/Q_s^2(x)$. The saturation scale is given
by $Q_s^2(x)\propto \Lambda_{\mathrm{QCD}}^2\, x^{-\lambda}$ and gives the
onset of nonlinear corrections to the QCD dynamics.  As the DVCS data are
predominantly at intermediate $Q^2$, we will be interested in small size
colour dipoles (here, the dipole cross section is dominated by colour 
transparency behavior). In this particular region, we found in Ref.
\cite{Favart:2003cu} the qualitative behavior for the total DVCS cross
section, taking into account only the non-forward kinematic effect,
\begin{eqnarray}
 \sigma(\gamma^* \,p\rightarrow \gamma \,p) \propto  
 \frac{Q_s^4}{Q^4}\,\left[ 1+\log \left( \frac{Q^2}{Q_s^2(x)} 
 \right)\,\right]^2 \,,
 \label{qualy}
\end{eqnarray}
where the large logarithm in Eq. (\ref{qualy}) originates from the intermediate
region $2/Q < \rrn < 2/Q_s(x)$ and the remaining comes from the region
$2/Q_s(x)<\rrn<2/Q$ in the dipole size integration in Eq.
(\ref{dvcsdip}).

The qualitative result above can be contrasted with the DIS case, where
the $\gamma^{*} p$ total cross section grows as $\sim W^{2\lambda}/Q^2$ at
intermediate $Q^2$. In particular, DIS data are proven to exhibit
a geometric scaling pattern in the scaling variable $\tau = Q^2/Q_s^2$. It
is also easy to demonstrate that DVCS data also should present such a
scaling behavior. In particular, at intermediate $Q^2$ it is shown that
the experimental DIS cross section behaves as
$\sigma(\gamma^*p\rightarrow X) \simeq 1/\tau=Q_s^2/Q^2$. Putting 
this information together and the result for the DVCS cross section 
presented in Eq.~(\ref{qualy}) into the definition of the skewing factor, Eq.
(\ref{R_def_ap}), one obtains,
\begin{eqnarray} 
  R\,(W,Q^2) \propto \left[ 1+ \log \left(
  \frac{Q^2}{Q_s^2(x)} \right)\,\right]^{-1} \,, \label{R_approx}
\end{eqnarray} 
where in the DESY-HERA domain the saturation scale takes values of
$Q_s^2=1\!-\!2$ GeV$^2$. Therefore, the qualitative $Q^2$ behavior for the
skewing factor takes the form $R\propto 1/(1+\log Q^2)$. On the other
hand, one has $Q_s^2\simeq x^{-\lambda} \approx (W/Q)^{2\lambda}$ which at
fixed virtualities the skewing factor behaves as $R\propto
1/(1-\varepsilon \log W)$. These qualitative results are in agreement with
the behavior of the experimental extraction of $R$. One notice that the
early theoretical determinations for $R$ predict $R \simeq
\log^{-1}[1+(Q^2/M_0^2)]$, with $M_0^2\simeq 0.4$-0.6 GeV$^2$, in close
proximity with the present estimate~\cite{Frankfurt:1997at}.

\section{Summary}

The skewing factor $R$, defined as the ration of the imaginary parts of
the DIS and DVCS amplitudes has been extracted from experimental data
for the first time.
In its determination one uses the recent Deeply Virtual
Compton Scattering (DVCS) and the inclusive inelastic cross section
measurements at DESY-HERA. The main theoretical uncertainties come from the
$b$ slope and from the estimate for the real part contribution. One
founds a $Q^2$ dependence qualitatively consistent with the form $R\propto
1/\log Q^2$. No $W$ dependence is observed within the current
precision. The mean value $R\simeq 0.5$ is consistent with previous
theoretical estimates in the jet aligned model and colour dipole picture. The
extracted $R$ is contrasted to theoretical predictions of NLO QCD and
colour dipole approaches. The first one uses two different GPD
parameterisations and produces a consistent description for the $W$ and $Q^2$
dependences of the skewing factor. The colour dipole formalism,
supplemented by a dipole cross section including QCD evolution and
off-diagonal corrections successfully describes $R$. In particular, a
qualitative understanding of the dependences of $R$ is obtained relying on
simple arguments in the dipole formalism. This study confirms a behavior
of the form $R\propto 1/(1+\log Q^2)$ and flat dependence upon $W$.
Beyond the successfully described dependences in $Q^2$ and $W$ achieved
here, to perform a further
discussion on the normalisation of $R$, a better agreement within 
experimental DVCS cross section measurements has to be achieved.

\section*{Acknowledgments}
The authors thank the support of the High Energy Physics
Group at Institute of Physics of Porto Alegre.
The work of L. Favart is supported by the FNRS of Belgium (convention
IISN 4.4502.01).

\end{document}